\documentclass[12pt]{article}
\usepackage{amsmath}
\usepackage{threeparttable}
\usepackage{multirow}
\usepackage{bbm}
\usepackage{ntheorem}
\usepackage{upgreek}
\usepackage{bigints}
\usepackage{colonequals}
\usepackage[official]{eurosym}
\usepackage[margin=1in]{geometry}
\usepackage{caption}
\usepackage{subcaption}
\usepackage{graphicx}
\usepackage[onehalfspacing]{setspace}
\usepackage{natbib}
\usepackage{booktabs}
\usepackage{comment}
\usepackage[compact]{titlesec}
\usepackage{xr}
\usepackage[colorlinks=true,citecolor=blue,urlcolor=blue]{hyperref}
\usepackage[utf8]{inputenc}

\renewcommand{\P}{\mathbbm{P}}

\newcommand{\I}{\mathbbm{1}}

\begin{document}
\newcommand{\blind}{0} 

\newcommand{\tit}{Natural Experiments}

\if0\blind

{\title{\tit\thanks{Prepared for ``Advances in Experimental Political Science'', edited by James Druckman and Donald Green, to be published by Cambridge University Press. I am grateful to Don Green and Jamie Druckman for their helpful feedback on multiple versions of this chapter, and to Marc Ratkovic and participants at the Advances in Experimental Political Science Conference held at Northwestern University in May 2019 for valuable comments and suggestions.  I am also indebted to Alberto Abadie, Matias Cattaneo, Angus Deaton, and Guido Imbens for their insightful comments and criticisms, which not only improved the manuscript but also gave me much to think about for the future.}}

\author{
Roc\'{i}o Titiunik \\ Professor of Politics\\ Princeton University
}

\date{\vspace{0.25in}\today  \\ \bigskip \bigskip }

\maketitle
\thispagestyle{empty}
}\fi

\if1\blind
\title{\bf \tit}
\maketitle
\thispagestyle{empty}
\fi
\onehalfspacing

\newpage 
\begin{abstract}
The term natural experiment is used inconsistently. In one interpretation, it refers to an experiment where a treatment is randomly assigned by someone other than the researcher. In another interpretation, it refers to a study in which there is no controlled random assignment, but treatment is assigned by some external factor in a way that loosely resembles a randomized experiment---often described as an ``as if random'' assignment. In yet another interpretation, it refers to any non-randomized study that compares a treatment to a control group, without any specific requirements on how the treatment is assigned. I introduce an alternative definition that seeks to clarify the integral features of natural experiments and at the same time distinguish them from randomized controlled experiments. I define a natural experiment as a research study where the treatment assignment mechanism (i) is neither designed nor implemented by the researcher, (ii) is unknown to the researcher, and (iii) is probabilistic by virtue of depending on an external factor. The main message of this definition is that the difference between a randomized controlled experiment and a natural experiment is not a matter of degree, but of essence, and thus conceptualizing a natural experiment as a research design akin to a randomized experiment is neither rigorous nor a useful guide to empirical analysis. Using my alternative definition, I discuss how a natural experiment differs from a traditional observational study, and offer practical recommendations for researchers who wish to use natural experiments to study causal effects.
\end{abstract}


\clearpage
\setcounter{page}{1}

The framework for the analysis and interpretation of randomized experiments is routinely employed to study interventions that are not experimentally assigned but nonetheless share some of the characteristics of randomized controlled trials. Research designs that study non-experimental interventions invoking tools and concepts from the analysis of randomized experiments are sometimes referred to as \textit{natural experiments}. However, the use of the term has been inconsistent both within and across disciplines.

My first goal is to introduce a definition of natural experiment that identifies its integral features and distinguishes it clearly from a randomized experiment where treatments are assigned according to a known randomization procedure that results in full knowledge of the probability of occurrence of each possible treatment allocation. I call such an experiment a randomized \textit{controlled} experiment, to emphasize that the way in which the randomness is introduced is controlled by the researcher and thus results in a known probability distribution. One of the main messages of the new definition is that the difference between a randomized controlled experiment and a natural experiment is not a matter of degree, but of essence, and therefore conceptualizing a natural experiment as a research design that approximates or is akin to a randomized experiment is neither rigorous nor a useful guide to empirical analysis.

I then consider the ways in which a natural experiment in the sense of the new definition differs from other kinds of non-experimental or observational studies. The central conclusions of this discussion are that, compared to traditional observational studies where there is no external source of treatment assignment, natural experiments (i) have the advantage of more clearly separating pre- from post-treatment periods and thus allow for a more rigorous falsification of its assumptions; and (ii) can offer an objective (though not directly testable) justification for an unconfoundedness assumption.

My discussion is inspired and influenced by the conceptual distinctions introduced by \cite{Deaton2010-JEL} in his critique of experimental and quasi-experimental methods in development economics \citep[see also][]{Deaton2020-Chapter}, and based on the potential outcomes framework developed by \cite{Neyman1923} and \cite{Rubin1974-JEducPsy}---see also \cite{Holland1986-JASA} for an influential review, and \cite{ImbensRubin2015-Book} for a comprehensive textbook.

The use of natural experiments in the social sciences was pioneered by labor economists around the early 1990s \citep[e.g.][]{Angrist1990-AER, AngristKrueger1991-QJE, CardKrueger1994-AER},  and has been subsequently used by hundreds of scholars in multiple disciplines, including political science. My goal is neither to give a comprehensive review of prior work based on natural experiments, nor a historical overview of the use of natural experiments in the social sciences. For this, I refer the reader to \cite{AngristKrueger2001-JEP}, \cite{Craig-etal2017-AnnRevPH}, \cite{Dunning2008-PRQ, Dunning2012-Book}, \cite{Meyer1995-JBES}, \cite{Petticrew-etal2005-PublicHealth}, \cite{RosenzweigWolpin2000-JEL}, and references therein. See also \cite{Abadie-Cattaneo_2018_ARE} for a recent review of program evaluation and causal inference methods.

\section{Two Examples}

I start by considering two empirical examples, both of which are described by their authors as natural experiments at least once in their manuscripts. The first example is the study by \cite{Lassen2005-AJPS}, who examines the decentralization of city government in Copenhagen, Denmark. In 1996, the city was divided into fifteen districts, and four of those districts were selected to introduce a local administration for four years. The four treated districts were selected from among the fifteen districts to be representative of the overall city in terms of various demographic and social characteristics. In 2000, a referendum was held in the entire city, giving voters the option to extend decentralization to all districts or eliminate it altogether. \citeauthor{Lassen2005-AJPS} compares the referendum results of treated versus control districts to estimate the effect of information on voter turnout. The assumption is that voters in the treated districts are better informed about decentralization than control voters; and the hypothesis tested is that uninformed voters are more likely to abstain from voting, which at the aggregate level should lead to an increase in voter turnout in the treated districts. \cite{Lassen2005-AJPS} considers the assignment of districts to the decentralization/no decentralization conditions as ``exogenously determined variation'' (p. 104) in whether city districts have first-hand experience with decentralization. \cite{Lassen2005-AJPS} then uses decentralization as an instrument for information, but I focus exclusively on the ``intention-to-treat'' effect of decentralization on turnout.

The second example is \cite{Titiunik2016-PSRM}, where I studied the effect of term length on legislative behavior in the state senates of Texas, Arkansas, and Illinois. In these states, state senators serve for four years and are staggered, with half of the seats up for election every two years. Senate districts are redrawn immediately after each decennial census to comply with the constitutionally mandated requirement that all districts have equal population. But state constitutions also mandate that all state senate seats must be up for election immediately after reapportionment. In order to comply with this requirement and keep seats staggered, in the first election after reapportionment all seats are up for election but the seats are randomly assigned to two term-length conditions: either serve 2-years immediately after the election (and then two consecutive 4-year terms) or serve 4-years immediately after the election (and then one 4-year term and another 2-year term). \cite{Titiunik2016-PSRM} used the random assignment to 2-year and 4-year terms that occurred after the 2002 election under newly redrawn districts to study the effect of shorter terms on abstention rates and bill introduction during the 2002-2003 legislative session. 

\section{Two common definitions of natural experiment}

The two examples presented above share a standard program evaluation setup \citep[e.g.][]{Abadie-Cattaneo_2018_ARE,ImbensWooldridge2009-JEL}, where the researcher is interested in studying the effect of a binary treatment or intervention (decentralization, short term length) on an outcome of interest (voter turnout, abstention rates). They also have in common that neither study was designed by the researcher: the rules that determined which city districts had temporary decentralized governments or which senators served two-year terms were decided, respectively, by the city government  of Copenhagen and the state governments of Arkansas, Illinois, and Texas---not by the researchers who published the studies. In both cases, the researcher saw an opportunity in the allocations of these interventions to answer a question of long-standing scientific and policy interest. 

Despite their similarities, the examples have one crucial difference. In the decentralization study by \cite{Lassen2005-AJPS}, the assignment of districts to the decentralized/not-decentralized conditions was not determined by a physical randomization device, but rather by officials seeking to select treated districts that were representative of the city as a whole. In contrast, the assignment of senate seats to two-year or four-year terms was based on a fixed-margins randomization device that gave each senator the same probability of serving two or four year terms.\footnote{See \cite{Titiunik2016-PSRM} for details on the assignments in each state. In Texas, for example, the 35 senate seats were allocated by creating 17 pieces of paper marked with a ``2'' and 18  marked with a ``4'', mixing all pieces inside a bowl, and having each of the 35 elected senators draw one piece of paper without looking.}

Common definitions of the term \textit{natural experiment} include the researcher's lack of control over the treatment assignment as an integral feature. At the same time, researchers who invoke natural experiments assume that, despite the lack of control over the assignment of the treatment, some external forces of nature imbue the design with some superior credibility for causal inference relative to other observational studies where such external factors are absent. But current definitions of natural experiments do not explicitly describe the source of such superior credibility, other than invoking an analogy between the `natural-experimental' treatment assignment and the kind of treatment assignment that governs randomized controlled trials. There are two ways in which this analogy is made, one is literal, and the other is figurative, leading to two different definitions of a natural experiment. 

In the literal interpretation of \citet[][p. 15]{GerberGreen2012-Book}, a natural experiment is a situation in which there is random assignment of a treatment via a randomization device, but this assignment is not under the control of the researcher. According to this definition, the term length study in \cite{Titiunik2016-PSRM} is a natural experiment, but the decentralization study in \cite{Lassen2005-AJPS} is not. Other examples that conform to this definition of natural experiment include \cite{EriksonStoker2011-APSR}, who use the Vietnam draft lottery to study the effect of the military draft on political attitudes, and \cite{Bhavnani2009-APSR}, who uses a rule that randomly reserves one third of seats to women candidates in India's local elections to study the impact of reservations on women's future electoral success.\footnote{ \citet[][p. 16]{GerberGreen2012-Book}, following \cite{CookCampbell1979-Book} and \cite{CookCampbellShadish2002-Book}, use the term \textit{quasi-experiment} to refer to studies such as \cite{Lassen2005-AJPS} where no actual randomization device is employed.}

This definition has the advantage of being precise. Understood as a randomized experiment controlled by an external party,  a natural experiment can be analyzed by directly applying the standard tools from the analysis of randomized experiments. To be sure, the interpretation of the estimated parameter can still pose serious challenges when the groups that the randomization deems comparable are not directly informative about the parameter of scientific interest \citep{SekhonTitiunik2012-APSR}.\footnote{\cite{SekhonTitiunik2012-APSR} consider various examples of natural experiments where this phenomenon occurs, including the study by \cite{Bhavnani2009-APSR} cited above.} But interpretation issues aside, the assumptions and methods for estimation and inference under controlled randomization are well established. Because this definition of natural experiment is conceptually clear and its implementation relatively uncontroversial, it is not the focus of my discussion.  

Instead, my interest lies in another, widely used definition that interprets a natural experiment as some sort of imperfect approximation to a randomized controlled experiment. According to this figurative definition, a natural experiment is a situation in which an external event introduces variation in the allocation of the treatment, and the researcher uses the external event as the basis to claim that the treatment is ``as good as random'' or ``as if random,'' but no physical randomization device is employed by any human being. 

Scholars who employ this notion of natural experiments do not typically offer a formal definition of ``as if randomness'', but rather refer heuristically to an analogy or comparison with randomized experiments. Different versions of this analogy have been offered in political science, economics, public health, and other sciences. In political science, \cite{Dunning2008-PRQ} defines a natural experiment as a study in which the data come ``from naturally occurring phenomena'' (p. 282) where the treatment is not assigned randomly but the researcher makes ``a credible claim that the assignment of the nonexperimental subjects to treatment and control conditions is `as if' random'' (p. 283). In economics, \cite{Meyer1995-JBES} defines a natural experiment as a study that investigates ``outcome measures for observations in treatment groups and comparison groups that are not randomly assigned'' (p. 151), and  \cite{AngristKrueger2001-JEP} as a situation ``where the 
forces of nature or government policy have conspired to produce an environment somewhat akin to a randomized experiment'' (p. 73). In public health, \cite{Petticrew-etal2005-PublicHealth} define natural experiments in contrast to randomized experiments, as designs in which ``the researcher cannot control or withhold the allocation of an intervention to particular areas or communities,  but where natural or predetermined variation in allocation occurs'' (p. 752).

This definition of a natural experiment, which I shall name the `as-if random' definition, seems to be more common among empirical researchers than the definition of \cite{GerberGreen2012-Book}. Most empirical researches who invoke natural experiments refer to cases where a treatment is allocated by forces outside their control and not based on a randomization device.

\section{Conceptual Distinctions}
\label{sec:concept}

Given the widespread use of the as if random interpretation of a natural experiment, my focus in this chapter is on research designs of this type. That is, I focus on research designs where there is no physical randomization device controlled by a human being, but there is some external factor that determines treatment assignment. However, I depart from the as-if random definition of natural experiments and instead present a definition in which natural experiments are defined \textit{in contrast} to randomized experiments as opposed to akin to them.  My definition encompasses the spirit of the as-if random understanding of a natural experiment but introduces a more rigorous understanding of the role of experimental manipulation and random assignment, introducing conceptual distinctions that have so far remained fused.  My discussion builds most directly on prior arguments by \cite{Deaton2010-JEL} and on several definitions discussed by \cite{ImbensRubin2015-Book}.

 The case of experimentation without randomization is beyond the scope of my discussion, but is worth considering at least briefly. Loosely, an experiment is a study in which the researcher executes a direct controlled intervention over some process in order to test a hypothesis and/or explore potential mechanisms. An experimental intervention need not be randomly assigned, and indeed non-random experiments are ubiquitous in the natural sciences where there is sufficient prior knowledge (such as established laws of physics) to plausibly create a controlled environment. 

Given the meaning of the term \textit{experiment}, the term \textit{natural experiment} seems to be an oxymoron, since the adjective \textit{natural} often refers to the researcher's lack of control over the treatment assignment. A controlled randomized experiment is thus a special case of an experiment, and the opposite of an experiment is an observational study (where the researcher is unable to experiment or control the conditions). In my proposed usage,  discussed at length below, a natural experiment is (oxymoronically) a special case of an observational study, not a special case of an experiment. Rather than changing established usage of these terms, in the following pages I seek to clarify the concepts that these terms refer to.

\subsection{Randomized Experiments} 

The first step to arrive at a precise and encompassing definition of a natural experiment requires that we define the term \textit{randomized experiment} with some precision. For this, I follow the Neyman-Rubin potential outcomes framework \citep{Neyman1923, Rubin1974-JEducPsy,Holland1986-JASA} and introduce standard notation. I assume that the researcher studies a population of $n$ units, indexed by $i=1,2,\ldots,n$, and her goal is to analyze the effect of a binary intervention or treatment $Z$, with $Z_i=1$ if $i$ is assigned to the treatment condition $Z_i=0$ if $i$ is assigned to the control. Each unit $i$ has two potential outcomes corresponding to each one of the treatment conditions, with $Y_i(1)$ the outcome that $i$ would attain under treatment and $Y_i(0)$ the outcome that $i$ would attain under control. The observed outcome is $Y_i = Z_i Y_i(1) + (1-Z_i) Y_i(0)$, and $\mathbf{X}_i$ is a vector of $k$ covariates determined before the treatment is assigned (hereafter called predetermined covariates). The individual-level variables are collected in the $n \times 1$ vectors (or $n \times k$ matrix), $\mathbf{Y(1)}$, $\mathbf{Y(0)}$, $\mathbf{X}$, and $\mathbf{Z}$. 

This notation  can be used to describe the two examples above. In the \cite{Lassen2005-AJPS} study, the units are city districts, $Z_i=1$ if a district's government is decentralized and $Z_i=0$ otherwise, and $Y_i$ is district-level voter turnout. In the \cite{Titiunik2016-PSRM} study, the units are state senators, $Z_i=1$ if senator $i$ serves a two-year term after redistricting and $Z_i=0$ if she serves a four-year term instead, and $Y_i$ is the abstention rate or number of bills introduced during the post-redistricting legislative session.

The vector $\mathbf{Z}=\mathbf{z}$ gives the particular arrangement of treated and control units that occurred. For example, if $n=5$ and $\mathbf{z} = c(1,0,0,1,1)$, units 1, 3, and 5 were assigned to the treatment group, and units 2 and 3 were assigned to the control. I define the assignment mechanism $\Pr(\mathbf{Z} | \mathbf{X}, \mathbf{Y(1)}, \mathbf{Y(0)})$ as in \cite{ImbensRubin2015-Book}. This function gives the probability of occurrence of each possible value of the treatment vector $\mathbf{Z}$. It therefore takes values in the interval $[0,1]$, and satisfies $\sum_{\mathbf{z} \in \{0,1\}^n} \Pr(\mathbf{z} | \mathbf{X}, \mathbf{Y}(0),\mathbf{Y}(1)) = 1$ for all $\mathbf{X}, \mathbf{Y}(0),\mathbf{Y}(1)$.  From this, we can define the unit-level assignment probability $p_i$ as the sum of the probabilities associated with all the assignments that result in unit $i$'s receiving the treatment, $p_i(\mathbf{X}, \mathbf{Y}(0),\mathbf{Y}(1))  = \sum_{\mathbf{z}: z_i=1} \Pr(\mathbf{z} | \mathbf{X}, \mathbf{Y}(0),\mathbf{Y}(1))$. 

\cite{ImbensRubin2015-Book} define randomized experiments in terms of restrictions placed on the assignment mechanism. I restate some of these restrictions and their definition of randomized experiment, which I then use as the basis of my discussion. 

The first restriction I consider requires that every unit be assigned to treatment with probability strictly between zero and one. Formally, $\Pr(\mathbf{Z} | \mathbf{X}, \mathbf{Y}(0),\mathbf{Y}(1))$ is a probabilistic assignment \cite[p.38]{ImbensRubin2015-Book} if

	\begin{equation}
	\label{eq:probass}
	0  < p_i(\mathbf{X}, \mathbf{Y}(0),\mathbf{Y}(1)) < 1 \;\;\;\; \text{for every $i$, for each $\mathbf{X}, \mathbf{Y}(0),\mathbf{Y}(1)$}\text{.} 
	\end{equation}

An assignment is probabilistic when every unit has both a positive probability of being assigned to the treatment condition and a positive probability of being assigned to the control condition---in other words,  when all units are ``at risk'' of being assigned to both conditions before the treatment is in fact assigned. Important for our purposes, a probabilistic assignment rules out deterministic situations where, conditional on $\mathbf{X}$, $\mathbf{Y}(0)$, and $\mathbf{Y}(1)$, units know with certainty which of the two treatment conditions they will be assigned to.

Given these possible restrictions on the assignment mechanism, \cite{ImbensRubin2015-Book} offer a definition of a randomized experiment. 

\noindent\textbf{Definition RE (Randomized Experiment, Imbens and Rubin 2015, p.40).} \textit{
A randomized experiment is a study in which the assignment mechanism satisfies the following properties:} 
\begin{enumerate}
	\item[(C)] $\Pr(\mathbf{Z} | \mathbf{X}, \mathbf{Y}(0),\mathbf{Y}(1))$  \textit{is controlled by the researcher and has a known functional form}.
	\item[(P)] $\Pr(\mathbf{Z} | \mathbf{X}, \mathbf{Y}(0),\mathbf{Y}(1))$  \textit{is probabilistic}.
\end{enumerate}
	
Several aspects of this definition are relevant for our purposes. First, the word `randomized' in the definition stems from condition (P) (probabilistic assignment), while the word `experiment' stems from condition (C) (researcher's knowledge and control). The researcher designs and controls the assignment of the treatment, thus creating an experiment or controlled manipulation, and this assignment is not deterministic, in the sense that no unit can rule out ex-ante the possibility of being assigned to one of the conditions.

None of the empirical examples introduced above satisfies this definition of a randomized experiment, but for somewhat different reasons. In \cite{Titiunik2016-PSRM}, the assignment mechanism is both probabilistic and known, but is not under the researcher's control and thus violates the control part of condition (C). In \cite{Lassen2005-AJPS}, both parts of condition (C) are violated, as the researcher has neither control over the assignment mechanism nor knowledge of its exact functional form. 

Second, this definition clearly separates the notion of randomization from the notion of ``valid'' comparison groups or lack of confounders, a distinction that is essential for characterizing natural experiments. Definition RE explicitly allows for the potential outcomes to affect the assignment mechanism, making clear that a probabilistic assignment does not guarantee that treated and control groups will be comparable, in the sense that it does not guarantee that the treatment is (conditionally) independent of the potential outcomes. Such an unconfoundedness condition must be added as a separate requirement. 

Formally, the assignment mechanism $\Pr(\mathbf{Z} | \mathbf{X}, \mathbf{Y}(0),\mathbf{Y}(1))$ is unconfounded \citep[p. 38]{ImbensRubin2015-Book} if it satisfies
\begin{equation}
\label{eq:unconass}
\Pr(\mathbf{Z} | \mathbf{X}, \mathbf{Y}(0),\mathbf{Y}(1)) = \Pr(\mathbf{Z} | \mathbf{X}, \mathbf{Y}(0)^\prime,\mathbf{Y}(1)^\prime) \;\;\; \text{for all} \;\;\; \mathbf{Z},\mathbf{X}, \mathbf{Y}(0),\mathbf{Y}(1),\mathbf{Y}(0)^\prime,\mathbf{Y}(1)^\prime.
\end{equation} 

An unconfounded assignment is one in which the probability of each possible treatment allocation vector is not a function of the potential outcomes. This property is violated when, for example, units who have higher potential outcomes are more likely to be assigned to the treatment condition than to the control even after conditioning on the available observable characteristics. In general, any study where units self-select into the treatment based on characteristics unobservable to the researcher that correlate with their potential outcomes constitutes an assignment mechanism that is not unconfounded. When a randomized experiment also satisfies unconfoundedness, \cite{ImbensRubin2015-Book} call it an unconfounded randomized experiment.

Building on the above definitions, I now state a definition of a randomized controlled experiment. (As I discuss below, this definition is different from \citeauthor{ImbensRubin2015-Book}'s definition of an unconfounded randomized experiment.)

\noindent\textbf{Definition RCE (Randomized Controlled Experiment)} \textit{A randomized controlled experiment (RCE) is a study in which the assignment mechanism satisfies the following properties:} 
	\begin{enumerate}
		\item[(D)] $\Pr(\mathbf{Z} | \mathbf{X}, \mathbf{Y}(0),\mathbf{Y}(1))$ \textit{is designed and implemented by the researcher}. 
		\item[(K)] $\Pr(\mathbf{Z} | \mathbf{X}, \mathbf{Y}(0),\mathbf{Y}(1))$ \textit{is known to the researcher}.		
		\item[(P)] $\Pr(\mathbf{Z} | \mathbf{X}, \mathbf{Y}(0),\mathbf{Y}(1))$ \textit{is probabilistic by means of a randomization device whose physical features ensure that $\Pr(\mathbf{Z} | \mathbf{X}, \mathbf{Y}(0),\mathbf{Y}(1))$ is unconfounded}. 
	\end{enumerate}

In a randomized controlled experiment (RCE) as I have defined it, the assignment mechanism is probabilistic, designed and implemented by the researcher, known to the researcher, and not a function of the potential outcomes (possibly after conditioning on observable characteristics). The latter condition means that the probability that the treatment assignment vector $\mathbf{Z}$ is equal to a given  $\mathbf{z}$ is entirely unrelated to the unit's potential outcomes, possibly after we have conditioned on $\mathbf{X}$. My definition of a RCE is similar to \citeauthor{ImbensRubin2015-Book}'s definition of an unconfounded randomized experiment, with one key difference. In the RCE definition, condition (P) explicitly requires that unconfoundedness be a direct consequence of the type of physical randomization device used to allocate the treatment probabilistically. This explicitly links unconfoundedness to the randomization device. 

The joint requirements of full control of the design and implementation (D), and knowledge (K) of the assignment mechanism imply that in a RCE, the treatment assignment mechanism is fully reproducible. In most cases, full knowledge and reproducibility of the assignment mechanism will be a direct consequence of the researcher's being in control of the treatment assignment, and thus condition (K) will be implied by condition (D). 

However, sometimes knowledge of the mechanism occurs despite the researcher not being in control of the experiment, which is why I separate conditions (D) and (K) in the definition. This occurs in experiments where, just as in a RCE, the treatment assignment is probabilistic and unconfounded by virtue of the use of a physical randomization device, but where the design and implementation of the assignment mechanism are not under the control of the researcher. I shall call this type of experiment a randomized third-party experiment (RTPE).\footnote{Others have called this a randomized policy experiment. See, for example, \cite{Clayton2015-CPS}.} I define it below for completeness.

\noindent\textbf{Definition RTPE (Randomized Third-Party Experiment)} \textit{A randomized third-party experiment (RTPE) is a study in which the assignment mechanism satisfies the following properties:}
	\begin{enumerate}	
	\item[(D$^\prime$)] $\Pr(\mathbf{Z} | \mathbf{X}, \mathbf{Y}(0),\mathbf{Y}(1))$  \textit{is designed and controlled by a third party}.

	\item[(K)] $\Pr(\mathbf{Z} | \mathbf{X}, \mathbf{Y}(0),\mathbf{Y}(1))$ \textit{is known to the researcher}.
				
	\item[(P)] $\Pr(\mathbf{Z} | \mathbf{X}, \mathbf{Y}(0),\mathbf{Y}(1))$  \textit{is probabilistic by means of a randomization device whose physical features ensure that $\Pr(\mathbf{Z} | \mathbf{X}, \mathbf{Y}(0),\mathbf{Y}(1))$ is unconfounded}. 
	\end{enumerate}

\subsection{Three senses of the word random}
Some of the ambiguity regarding natural experiments has stemmed from the failure to properly distinguish randomness from unconfoundedness, and mistakenly assuming that randomness in the assignment mechanism automatically guarantees treatment and control groups that are comparable in all relevant respects. At least part of the ambiguity seems to stem from the different senses of the word random, which are used sometimes interchangeably to describe both natural experiments and randomized controlled experiments. 

I now discuss different meanings of random and their relationship to unconfoundedness, relying on a related distinction between externality and exogeneity introduced by \cite{Deaton2010-JEL} in his critique of the use of natural experiments as sources of instrumental variables (IV). Following a terminology first adopted by Heckman, \citeauthor{Deaton2010-JEL} distinguishes between an instrument being ``external'' to refer to variables that are determined outside the system and ``exogenous'' to refer to the orthogonality condition that is needed for consistent estimation of the parameter of interest in an IV context. My focus in this chapter is on studies where interest lies directly on the effect of $Z$ on $Y$ and not on its effect via another variable, so I ignore concerns about the exclusion restriction.  However, I will show that even in this simpler case, the distinction between the externality of $Z$ and the type of ``randomization'' that such externality creates is essential to understand the ways in which natural experiments differ from randomized controlled experiments. 

It is well known that randomized controlled experiments fail to guarantee the IV exclusion restriction \citep{AngristImbensRubin1996-JASA}. Thus, in IV settings, natural experiments and RCEs are on a more equal footing, in the sense that neither can guarantee the assumptions to identify the treatment effect of interest. When it comes to the ``reduced form'' effect of $Z$ on $Y$, however, natural experiments face unique challenges that are absent in RCEs. My interest in this section is to discuss these particular challenges, and for this reason I focus on the ``reduced form'' effect of $Z$ on $Y$. However, my discussion also applies to IV settings, because the challenges faced by natural experiments in identifying the reduced form effect remain when natural experiments are used as a ``source of instruments'' \citep[][p. 73]{AngristKrueger2001-JEP}.

I consider three different uses of the term ``random'', all of which have been used to characterize natural experiments---though I do not mean to imply that these are the only three ways in which the term random has been used in the history of science.  The first is what I call the \textit{colloquial} definition of random; this is the first sense listed by the \citeauthor{MerriamWebster-Online} dictionary, which defines the adjective random as ``lacking a definite plan, purpose, or pattern'' and further clarifies that this use ``stresses lack of definite aim, fixed goal, or regular procedure.'' Used in this sense, a random treatment assignment refers to an assignment mechanism that follows an arbitrary, inscrutable plan that has no clear pattern. 

The notion of inscrutability behind the colloquial sense of random is similar to the concept of Knightian uncertainty in economics. In his seminal study, \cite{Knight1921-Book} used the term \textit{risk} to refer to the kind of uncertainty that is measurable and quantifiable with objective probabilities, and reserved the term \textit{uncertainty} to refer to situations where the randomness cannot be objectively quantified and thus cannot be insured in the market. A similar distinction was  advanced by \cite{Keynes1921-Book}, see the discussion in \cite{LeroySingell1987-JPE}.

The second meaning of the word random is most likely found in statistics textbooks. This sense of random, which I call the \textit{statistical} definition of random, refers to situations in which we have uncertainty about what event will occur, but we can precisely characterize all possible events that may occur and exactly quantify the probability with which each event will occur (analogous to Knightian risk). In this sense, a random treatment assignment is an assignment of units to treatment and control conditions in which the uncertainty can be completely and exactly quantified via the function $\Pr(\mathbf{Z} | \mathbf{X}, \mathbf{Y}(0),\mathbf{Y}(1))$, which specifies the probability of occurrence of each possible treatment allocation.  
Used in the statistical sense, randomization thus refers to ``the selection of an element $a$, from a set $A$, according to some probability distribution $P$ on $A$'' \citep{Berger1990-Chapter}. 

In his treatise on experimental design, \cite{Fisher1935-book} explicitly rejects the colloquial sense of random in his definition of a randomized experiment. While discussing an agricultural experiment that assigns land plots to various crops to test the relative yield of each crop variety, Fisher is explicit in ruling out haphazardness or arbitrariness. 

\begin{quote}
	``In each block, the five plots are assigned one to each of the five varieties under the test, and this assignment is made at random. This does not mean that the experimenter writes down the names of the varieties, or letters standing for them, in any order that may occur to him, but that he carries out a physical experimental process of randomisation, using means which shall ensure that each variety has an equal chance of being tested on any particular plot of ground.'' \citep[][p.51]{Fisher1935-book}
\end{quote}

The above passage suggests yet a third sense of random, which is in fact a particular case of the statistical definition.  This third definition equates randomness with a situation in which all possible outcomes are equally likely. This is the sense used by Fisher in the passage above, and even more explicitly described in \cite{Fisher1956-book} when he discusses random throws of a die:
\begin{quote}
	``(...) we may think of a particular throw, or of a succession of throws, as a \textit{random} sample from the aggregate, which is in this sense subjectively homogeneous and without recognizable stratification.'' \cite[][p. 35, emphasis in the original]{Fisher1956-book}
\end{quote}

When used in this third sense, a random assignment mechanism refers to a mechanism that gives every single possible arrangement of treated and control units the same probability of occurrence. For example, if an assignment mechanism allocates exactly $n_t$ units to treatment and $n-n_t$ units to control, it is random in this sense if $\Pr(\mathbf{Z}  = \mathbf{z}| \mathbf{X}, \mathbf{Y}(0),\mathbf{Y}(1)) = 1/\binom{n}{n_t}$ for all $\mathbf{z}$. I call this the \textit{equiprobable} sense of random.

In sum, the three senses of random refer to three different kinds of uncertainty. The colloquial sense means uncertainty that is arbitrary and inscrutable, not amenable to characterization by a clear pattern. The statistical sense of random refers to uncertainty that can be precisely characterized by a known probability distribution. And the equiprobable sense of random is a particular case of the statistical sense, and refers to uncertainty that is characterized by a known probability distribution that assigns equal probability to each possible outcome.

The ambiguous and overlapping usage of the term random is why defining a natural experiment as having an `as if' random treatment assignment lacks statistical rigor. If random is used in the colloquial sense, then the `as if' qualifier is not needed and distorts meaning, as random in the colloquial sense already refers to an arbitrary/inscrutable assignment. If random is used in the statistical sense, the  `as if' qualifier is simply incorrect. The assignment vector $\mathbf{Z}$  is a random variable, and as such it has some distribution over the sample space of assignments. Used in the statistical sense, a natural experiment has an exact random assignment, not an  `as if'  random assignment. Finally, used in the equiprobable sense, a natural experiment is typically not random at all: earthquakes are more likely to destroy huts than concrete buildings, rain on election day is more likely in Seattle than in Arizona, and abortion restriction laws are more likely to be passed in conservative than liberal states.

\subsection{Random assignment does not imply probabilistic assignment}

An assignment mechanism that is random in either the statistical or the equiprobable sense need not be probabilistic in the sense of equation (\ref{eq:probass}). For example, neither the statistical nor the equiprobable definition of random rules out a treatment assignment mechanism in which all units are assigned to treatment with probability one. This point is trivially true---a constant is a special case of a random variable in which all the probability mass is accumulated at a single value---but it matters for our purposes. Of course, since random assignment is usually discussed in the context of evaluating the effects of receiving a treatment relative to not receiving it, the existence of a comparison group in this context is presupposed. This is why Fisher does not explicitly include probabilistic in his definition of random, but it is clear that he does so implicitly. Informally, the requirement that the assignment be probabilistic is essential if our purpose is to obtain comparable treated and control units; otherwise, the treatment assignment may be perfectly correlated with confounders.\footnote{See \cite{HeckmanIchimuraTodd1998-ECMA} for a formal characterization of the bias introduced by violations of a probabilistic assignment in the context of selection on observables, which also applies immediately to stratified randomized experiments. If the assignment is deterministic for units with certain characteristics $\mathbf{X}=\mathbf{x}$, this introduces a lack of common support that impedes obtaining valid causal effects even if the assignment is unconfounded.}

The colloquial sense of random does rule out the particular deterministic assignment that assigns every unit to treatment (or to control), since in this case a very clear pattern of assignment would be discernible. However, other forms of non-probabilistic assignments are still compatible with the colloquial notion of randomness. For example, Fisher's farmer could decide that plots on the edge of the property line will always be assigned to the same crop. This decision would be entirely arbitrary, thus satisfying the colloquial definition of random. Moreover, to the external observer, this non-probabilistic assignment would be hard to catch, unless she happens to measure the proportion of boundary plots in treatment versus control groups. This point turns out to be important: in natural experiments, since the assignment mechanism is unknown to the researcher, she will not be able to distinguish probabilistic from deterministic assignments, because the assignment could be deterministic conditional on a characteristic that is unobserved to the researcher---which would give the appearance of a probabilistic assignment. 

A probabilistic assignment is therefore an assignment that is random in the statistical sense, with the added restriction that the probability distribution that characterizes the randomness not assign extreme (i.e., zero or one) individual probabilities.

\subsection{Random Assignment Does not Imply Unconfoundedness}
In a randomized experiment as stated in Definition RE, no unit has perfect control over which treatment it receives, in the sense that all units have ex-ante probability of being assigned to both the treated and control conditions. The assignment is therefore random in the statistical sense, governed by  $\Pr(\mathbf{Z} | \mathbf{X}, \mathbf{Y}(0),\mathbf{Y}(1))$. However, a probabilistic assignment does not imply an unconfounded assignment.

This point is easy to see in terms of our decentralization example. Imagine that in the \cite{Lassen2005-AJPS} study, some city districts have high crime, and reducing crime is the top priority of government administrators. Imagine also that decentralization gives districts more precise tools to combat and reduce crime. To say that the assignment is probabilistic or ``randomized'' is to say, for example, that districts lack the ability to perfectly and precisely self-select into the decentralization treatment which they believe will result in the most effective crime reduction. But this does not mean that high-crime districts have the same probability of being decentralized than low-crime districts. Perhaps officials from high-crime areas forcefully express their strong preference for decentralization to city administrators, and this results in their having a larger probability of receiving the treatment than low-crime areas. A probabilistic assignment only means that this probability is not one (nor zero); it does not mean that different types of units have the same probability of receiving treatment. If assignments with decentralized high-crime areas are more likely than assignments with decentralized low-crime areas, a naive, unadjusted comparison of treated versus control outcomes will not yield a valid estimate of the average effect of decentralization. A valid comparison requires that we reweight or stratify the observations based on the different probabilities of receiving treatment, something that is easy to do if we know the exact functional form of $\Pr(\mathbf{Z} | \mathbf{X}, \mathbf{Y}(0),\mathbf{Y}(1))$ but entirely unfeasible if this assignment mechanism is unknown.

One way to think of a confounded assignment is to see it as a blocked randomized experiment in which different ``types'' of individuals defined by potential outcomes are assigned to treatment with different probabilities. For example, imagine that all units have the same potential outcome under control, $Y_i(0)=y_0$ for all $i$. Defining high-types as units with $Y_i(1)-y_0 >0$ and low-types as units with $Y_i(1)-y_0 \leq 0$, we can conceive of a randomized experiment that violates unconfoundeness as a blocked randomized experiment where high-types are assigned to treatment with higher probability than low-types, and types are unobservable to the researcher. It is well known that the proper analysis of a stratified randomized experiment with treatment assignment probabilities that vary by strata requires accounting for the different strata, which in turn requires knowing the strata to which every unit belongs \citep[see, e.g.,][]{AtheyImbens2017-HandbookFE,GerberGreen2012-Book, ImbensRubin2015-Book}. In this example, failing to account for the different strata would over-estimate the true average treatment effect. In general, obtaining valid conclusions from an unconfounded block-randomized experiment is not feasible when the strata remain hidden from the researcher. In other words, chance does not imply comparability.

Finally, note that randomness in the equiprobable sense \textit{does} imply unconfoundedness. An assignment mechanism that is equiprobable is also unconfounded: any assignment that gives each vector $\mathbf{z}$ the same probability of being chosen is by construction attaching a constant probability to each $\mathbf{z}$, which as a consequence cannot be a function of the potential outcomes. But the converse is not true: an unconfounded assignment mechanism does not imply that each possible treatment assignment vector $\mathbf{z}$  must be equally likely. For example, a mechanism that uses a random device to allocate 2/3 of women and 1/3 of men to treatment is unconfounded, but is not random in the equiprobable sense when all units are considered as a whole (though it is equiprobable within gender blocks). An equiprobable random assignment mechanism is perhaps the simplest way to ensure an unconfounded assignment, which may be why the term ``random assignment'' is often used as a synonym for unconfoundedness.

\subsection{Physical Devices that Ensure Unconfoundedness}

Because a probabilistic assignment mechanism does not imply that the mechanism is unconfounded, it follows that the superior credibility of RCEs does not stem exclusively from random chance. Although chance or uncertainty are needed to ensure condition (P) in the RCE definition, chance alone is not enough to bestow an experiment with the ability to identify causal effects. Somewhat counter-intuitively, part of the power of randomized experimentation lies not in the creation of uncertainty, but rather in the use of physical randomization devices that are capable of assigning treatments without being influenced by the units' potential outcomes. These devices are more than the means by which the end of probabilistic assignment is achieved; they ensure that chance is introduced in a way that ensures identification of causal effects and the quantification of uncertainty. Without a physical randomization device that ensures knowledge of the probability distribution of the assignment mechanism, chance is not necessarily helpful.

To see this, consider the following strategies to introduce a probabilistic treatment assignment. We could stack paper applications on a desk and blow a fan at them, and then assign to treatment the applications that fall to the floor. Or we could have an octopus select applications,\footnote{For an octopus-based assignment mechanism, see the case of Paul the psychic of Oberhausen, e.g. \url{https://www.bbc.com/news/10420131}.} or let Fisher's farmer choose the applications ``in any order that may occur to him''. All of these strategies would be random in the colloquial sense. It might also be plausible to assume that all of these strategies would lead to a probabilistic assignment in the sense that, a priori, all applications would have a nonzero chance of being selected for treatment and for control---though this may be difficult to verify. However, it would be premature to claim that the assignment is unconfounded. For example, if the original pile of applications on the table were sorted alphabetically with Z at the bottom and A at the top, and the wind was more likely to blow away top applications, we would have more names in the A-L part of the alphabet assigned to treatment than to control. Since, for example, ethnicity often correlates strongly with last name, our treatment and control groups would very likely differ on ethnicity, and as a result on any other observable and unobservable that may correlate with it, such as immigration status, parent's education, neighborhood of residence, etc. Of course, in this case we would also have a hard time figuring out the exact functional form of $\Pr(\mathbf{Z} | \mathbf{X}, \mathbf{Y}(0),\mathbf{Y}(1))$. 

The randomness introduced by a physical randomization device is crucially different from the randomness of a fan or an octopus. By a physical randomization device, I mean a device and a set of rules that allow a researcher to introduce randomness  with a known probability distribution function---that is, in a controlled way. 

Examples of such physical randomization devices are varied. \cite{Fisher1935-book} describes a device based on a deck of cards for an agricultural experiment in which five plots of land are to be assigned randomly to five fertilizer varieties. Cards are numbered from 1 to 100 and repeatedly shuffled so that they are arranged in random order; the five treatments are numbered 1 through 5; and the experimenter draws one card for every plot. The fertilizer assigned to the plot is the remainder obtained when the number on the drawn card is divided by 5 if the number is not a multiple of 5; if it is, the plot is assigned to fertilizer 5. This procedure guarantees that the each fertilizer variety corresponds to 20 cards; since there 100 cards, the probability that each plot is assigned to each of the fertilizer varieties is 1/5.

Another randomization device is a rotating lottery drum where the researcher deposits balls or tickets containing numbers representing each of the experimental units. The balls or tickets are drawn after rotating the drum, ensuring that at any point each of the remaining balls has the same probability of being selected. This procedure was used, for example, to assign each one of the integers between 1 and 366 to each one of the possible birth dates in a year (including February 29) to select who would be drafted to the Vietnam War, the numbers  1 through 366 indicating the order in which men would be drafted. (The Vietnam lottery, however, seems to have failed to produce equally likely outcomes; see discussion below.)

In scientific studies conducted today, the most common mechanism to allocate treatments randomly is based on computer-generated pseudo-random numbers.  The principles underlying the generation of pseudo-random numbers offer important lessons for our discussion. Pseudo-random numbers can be generated in multiple ways, but all of them share the characteristic of being entirely predictable, directly ruling out the colloquial definition of random. For example, the Lehmer linear congruential algorithm \citep{Lehmer1951-ACLH,ParkMiller1988-ACM} requires the choice of a prime modulus $m$, an integer $a \in {2,3,\ldots, m-1}$,  and an initial value $x_0$. The value $x_1$ is generated as $x_1 = ak$ where $k\equiv x_0 \mod m$ is the remainder when $x_0$ is divided by $m$; and all subsequent values are generated as $x_{i+1} = a x_i \mod m$. Given the initial value $x_0$, the entire sequence is entirely determined, which illustrates the fundamental distinction between the colloquial and the statistical definition of random, lucidly summarized by \citeauthor{ParkMiller1988-ACM}:

\begin{quote}
	Over the years many programmers have unwittingly demonstrated that it is all too easy to `hack' a procedure that will produce a strange looking, apparently unpredictable sequence of numbers. It is fundamentally more difficult, however, to write quality software which produces what is really desired---a virtually infinite sequence of statistically independent random numbers, uniformly distributed between 0 and 1. This is a key point: strange and unpredictable is not necessarily random. \citep[][p.1193]{ParkMiller1988-ACM} 
\end{quote}

Formally demonstrating that randomization devices do in fact produce an equidistributed sequence of numbers is difficult, both because the physical properties of certain randomization devices can be complex \citep[e.g.][]{AldousDiaconis1986-AMM} and because demonstrating (and even defining) the randomness of a sequence is a hard mathematical problem \citep[see,e.g.,][]{DowneyHirschfeldt2010-Book,PincusSinger1996-PNAS,PincusKalman1997-PNAS}. Nonetheless, with our current knowledge of mathematics and algorithmic randomness, several randomization devices such as pseudo-random number generators or sufficiently shuffled cards are in fact able to produce independent, uniformly distributed numbers. I refer to such devices as \textit{proper} randomization devices, to distinguish them from randomization devices that appear but ultimately fail to produce equidistributed sequences.

The feature that proper randomization devices have in common is that (i) the allocation of units to the treatment/control conditions that they produce are entirely determined by their physical and statistical properties, which are by construction unrelated to the units' potential outcomes and thus result in an  unconfounded assignment mechanism; and (ii) these properties are known and well understood, which in turn implies that the assignment mechanism is entirely known and thus reproducible. Thus, proper physical randomization devices not only ensure that there is an element of chance in which unit receives treatment but also, by their very properties, they simultaneously guarantee that the assignment mechanism is unconfounded (and known). 

The use of a proper physical randomization device is as fundamental in its role to ensure unconfoundedeness as in its role to ensure random chance. We could introduce chance in treatment assignment using fans, octopus, or earthquakes. But only a fully known and reproducible physical randomization device can guarantee that this randomness can be used as the basis for inference and identification.

This guarantee, however, is not bullet proof. There are numerous and notable examples where the physical properties of randomization devices failed to produce unconfounded assignments because they were mistakenly believed to be proper devices. For example, the implementation of the 1970 Vietnam lottery is believed to have been defective (the capsules not sufficiently mixed), assigning systematically lower numbers to birth dates in later months, contrary to the uniform distribution that the lottery drum was supposed to produce \cite[see, e.g.,][]{Fienberg1971-Science}. This is a case where the physical properties of the device were mistakenly believed to produce an equiprobable assignment. For another example, see the Lanarkshire milk experiment \citep{Student1931-Biometrika}.

Such ``failures of randomization'' can invalidate a RCE, unless the true probabilities induced by the defective randomization device can be learned or discovered. However, note that in the case of the Vietnam lottery, researchers were able to detect the departure from an equiprobable assignment precisely because they believed that the physical randomization device guaranteed such an assignment, and because an equiprobable assignment has objective empirical implications (similar number of observations per birth month, etc.). This ability to detect departures from a known randomization distribution is only possible when such distribution can be specified ex-ante. It is precisely because we believe that the Vietnam lottery drums should have produced a uniform assignment that we discover that something must have been wrong with the device (or with our beliefs about the device). 

In contrast, in natural experiments, because we fundamentally ignore the distribution of the external assignment mechanism, we have no way of using the observed assignment to validate our beliefs about the physical randomization device used by nature.
 
\section{An Alternative Definition of Natural Experiment }
\label{sec:newdef}

The key feature of the as-if random interpretation of a natural experiment is the existence of an external factor or phenomenon that governs the allocation of treatment among units. This external phenomenon is most commonly ruled by the laws of nature (earthquakes, hurricanes, etc.) or the laws of government (minimum age restrictions, voting rules, etc.), and results in a treatment assignment that has been variously described as haphazard \citep{Rosenbaum2002-Book}, as-if random \citep{Dunning2008-PRQ}, naturally occurring \citep{Rutter2007-PPS}, not according to any particular order \citep{GouldLavyPaserman2004-QJE},  serendipitous \citep{Dinardo2016-Palgrave, RosenzweigWolpin2000-JEL}, unanticipated \citep{CarboneHallstromSmith2006-ERE}, unpredictable \citep{Dunning2012-Book}, unplanned \citep{LaliveZweimuller2009-RES}, quasi-random \citep{FuchsSchundelnHassan2016-HbMacro}, or a shock \citep{MiguelSatyanathSergenti2004-JPE}.

As discussed above, an arbitrary and unpredictable treatment assignment implies neither unconfoundedness nor knowledge (and thus reproducibility) of the assignment mechanism---two distinctive features of randomized controlled experiments. For this reason, I introduce a definition of a natural experiment that preserves the externality of the treatment assignment mechanism but, in contrast to prior interpretations, emphasizes its non-experimental qualities rather than its ``as-if randomness''. In my definition, the external phenomenon that governs treatment assignment ensures (in successful cases) that the assignment mechanism is probabilistic, but not that it is unconfounded.

 I first distinguish randomized controlled experiments from observational studies, and then define a natural experiment as a particular case of an observational study. For this, I consider two dimensions. The first is whether the researcher is in control of the design and implementation of the experiment. The second dimension is whether the probabilities associated with each possible treatment allocation are known. The four possible combinations of these two criteria are illustrated in Table \ref{tab:typology}.

\begin{table}[h]%
	\begin{center}	
		\caption{Typology of randomized experiments and observational studies}
		\label{tab:typology}
\scalebox{0.9}{			
		\begin{tabular}{llcc}
			& & \multicolumn{2}{c}{\textbf{Probabilities known to researcher}}\\
			& & Yes & No\\
			\cmidrule[\heavyrulewidth]{3-4}
			\multirow{1}{*}{\textbf{Designed \& implemented}} & Yes &  Randomized controlled experiment (RCE) & \multirow{2}{*}{Observational study}  \\
		    \hfill \textbf{by the researcher} 	& No & Randomized third-party experiment (RTPE) & \\
			\cmidrule[\heavyrulewidth]{3-4}
		\end{tabular}
}	
	\end{center}
\end{table}

Given a probabilistic treatment assignment, the difference between a randomized experiment and a non-experimental design depends crucially on both knowledge and control of the assignment mechanism. When a researcher controls the design and implementation of a probabilistic treatment assignment, she has full knowledge of all the probabilities associated with each possible treatment allocation. As a consequence, the randomization procedure is fully known and reproducible. This combination, represented by the top-left corner of Table \ref{tab:typology}, corresponds to RCEs as defined above. The rows of the table correspond exactly with condition (D) in the definition of a RCE. Condition (P) in the definition is satisfied implicitly if we assume that when a researcher designs and controls the assignment, she chooses a probabilistic assignment.\footnote{If her assignment has $p_i \in \{0,1\}$ for some units, the parameter of interest can always be redefined for those units whose probabilities are neither zero nor one.} And the unconfoundedness assumption (U) is implied by the assumption that the treatment allocation probabilities are fully known.\footnote{For example, if a researcher uses a higher probability of treatment assignment for patients who are known to benefit the most from treatment, this would appear to violate the unconfoundedness assumption. However, since we are assuming that the  researcher designed the experiment, she would know and be able to reproduce all treatment assignment probabilities for every unit, thus making the high/low potential benefit strata fully observable, which would restore unconfoundedness (conditional on potential benefit). Even if all units are assigned to treatment with a different probability and there are no strata, knowing these probabilities is sufficient to consistently estimate the average treatment effect, and perform exact Fisherian inference based on the sharp null hypothesis. As long as all probabilities are fully known, the possibility of violating unconfoundedness does not arise, or is inconsequential.}

Being in charge of the design and implementation of the randomized experiment, however, is a not a necessary condition to having full knowledge of the assignment mechanism. Researchers often discover randomized experiments that are designed and implemented by third parties such as policy makers. In some cases, the third party is willing to disclose all details regarding the assignment mechanism, and as a consequence all probabilities of treatment assignment become known to the researcher despite her not being in direct control of the experiment.  In Definition RTPE, I called this a randomized third-party experiment (RTPE). RTPEs  belong in the bottom-left cell of Table \ref{tab:typology}, where probabilities are known but the experiment is either not designed or not implemented by the researcher, or possibly both. If the treatment assignment mechanism is known, then even when the researcher is not in control of the assignment as in  \cite{Titiunik2016-PSRM}, well-defined treatment effects are identifiable, inference methods for the analysis of randomized experiments are ensured to be valid, and assumptions are falsifiable. 

Regardless of who designs and implements the experiment, if the probabilities associated with each possible treatment allocation are unknown to the researcher, the design is non-experimental---also known as an ``observational study''. My definition of an observational study follows \cite{ImbensRubin2015-Book}, who define it as a study in which ``the functional form of the assignment mechanism is unknown'' (p. 41). In contrast to a RTPE, where the lack of direct design or implementation is accompanied by knowledge of the probability occurrence of each treatment allocation, in an observational study the researcher fundamentally ignores or has no access to these probabilities.

In practice, cases that belong to the top-right of Table \ref{tab:typology} cell are rare because  a randomized experiment that is designed and implemented by the researcher typically implies that the treatment assignment mechanism is fully known to the researcher. However, there might be cases where the researcher controls the treatment assignment, but either the design or the implementation is faulty and as a consequence the exact treatment allocation probabilities are unknown---examples include the Vietnam lottery and the Lanarkshire milk experiment mentioned above. 

Given the above distinctions, I now introduce a new definition of natural experiment. 

\noindent\textbf{Definition NE (Natural Experiment)} \textit{A natural experiment is a study in which the assignment mechanism satisfies the following properties:}
\begin{enumerate}
	\item[($\widetilde{D}$)] $\Pr(\mathbf{Z} | \mathbf{X}, \mathbf{Y}(0),\mathbf{Y}(1))$ \textit{is neither designed nor implemented by the researcher}.

	\item[($\widetilde{K}$)] $\Pr(\mathbf{Z} | \mathbf{X}, \mathbf{Y}(0),\mathbf{Y}(1))$  \textit{is unknown to the researcher}.		
	
	\item[($\widetilde{P}$)] $\Pr(\mathbf{Z} | \mathbf{X}, \mathbf{Y}(0),\mathbf{Y}(1))$  \textit{is probabilistic by virtue of an external event or intervention that is outside the experimental units' direct control}.
\end{enumerate}

This definition is intentionally analogous to my prior definitions of a RCE and a RTPE, to facilitate a comparison. A natural experiment is a research design where the researcher is neither in charge of the design of the treatment assignment mechanism nor of its implementation (condition $\widetilde{D}$). Moreover, the treatment assignment mechanism is unknown (condition $\widetilde{K}$), which means that the researcher does not know and has no way of knowing the probabilities associated with each possible treatment allocation. The latter condition---assignment mechanism unknown---immediately implies that a natural experiment is an observational study. 

The third and last condition in the definition  ($\widetilde{P}$) captures what has often been invoked as the main feature of a natural experiment: its unpredictability as a result of the assignment mechanism's dependence on an external factor. A natural experiment is a special kind of observational study where the mechanism that allocates treatment is known to depend on an external factor. In my definition, this external factor is assumed to be the source of randomness that results in a probabilistic assignment mechanism, and thus captures the unpredictable component that has been emphasized in prior characterizations of natural experiments. 

Note that condition $\widetilde{P}$ is not directly verifiable or falsifiable. Although the existence of the external factor will typically be immediately verifiable, verifying that this external factor resulted in a probabilistic assignment will be considerably more difficult and often impossible. Thus, classifying an observational study as a natural experiment will require \textit{assuming} that the external forces of nature that intervened in the assignment of treatment did so in such a way as to produce a probabilistic assignment. The justification of this assumption will often rest on the argument that the experimental units have no ability to directly control the external factor, and thus have no ability to choose their treatment condition deterministically. This is a heuristic rather than a formal argument, as the units' lack of control of their own assignment is not by itself sufficient to ensure a probabilistic assignment---rather, the lack of control introduced by the external factor is simply used as the basis for \textit{assuming} that the assignment was governed, at least partly, by chance.

In a standard observational study, it is often impossible for the researcher to know which, if any, of the units that actually took the treatment were ex-ante at risk of not taking it.  In contrast, in a natural experiment, there is an external factor that serves as the basis for making such assumption. Although  the probability of receiving treatment is still possibly a function of potential outcomes, it is also affected by an external factor over which the units have no precise control. For example, even though families can choose to invest on more durable construction materials to protect against earthquakes or floods, the severity of natural disasters is not under any family's control and thus it is impossible for a family to precisely and perfectly guarantee that their house will not be destroyed by a natural disaster, which introduces an element of chance to which houses are in fact destroyed. The distinction is similar to that introduced by \cite{Lee_2008_JoE} between ``systematic or predictable components that can depend on individuals' attributes and/or actions'' and a ``random chance component'' that is uncontrollable from the point of view of the unit \citep[p. 681]{Lee_2008_JoE}.

Crucially, the externality is not absolute, but relative to the units who are receiving the treatment. This external factor implies that the units that are assigned to treatment or control lack precise control over the treatment condition they will receive, and thus that the treatment assignment mechanism is not fully under the control of the units who are the subjects of the study. Thus, external  means ``external to units'', not necessarily to other actors. 

For example, in the \cite{Lassen2005-AJPS} study, the assignment of districts to the decentralization condition depended on various factors. Some of those factors are units' characteristics such as population size and suburban status. These are examples of characteristics $\mathbf{X}$ that may be correlated with the units' potential outcomes and determined before the treatment is assigned. But \citeauthor{Lassen2005-AJPS}'s account of the decision process that governed the decentralization policy suggests that, despite their different characteristics, \textit{all} the districts in the sample were at risk of being assigned to the decentralization group. The assumption of probabilistic assignment is supported by the policymakers' account of how the decentralization policy was carried out.

However, even if condition ($\tilde{P}$) holds and the assignment is in fact probabilistic by virtue of the external factor, there remains a crucial obstacle. The central distinction between a RCE or RTPE and a natural experiment as I have defined it is that, in a natural experiment, the exact probabilities with which each possible treatment allocation could have occurred are fundamentally unknown. Thus, even if the external factor prevents the experimental units from having precise control over which treatment condition they receive, the researcher has fundamental uncertainty about the actual probabilities associated with each allocation. Thus, a research design that satisfies Definition NE is still insufficient to identify or make inferences about causal effects, and researchers need to invoke additional assumptions. I elaborate on this issue in the following two sections, after discussing the particular case of the regression discontinuity design. 

\subsection*{Is the Regression Discontinuity Design a Natural Experiment?}

I now discuss whether definition NE applies to the regression discontinuity (RD) design, a research design that has become widely used throughout the social and behavioral sciences (for an overview see \citeauthor{Cattaneo-Idrobo-Titiunik-Book-2019}, \citeyear{Cattaneo-Idrobo-Titiunik-Book-2019} and \citeauthor{Cattaneo-Titiunik-VazquezBare_2020_SAGE} \citeyear{Cattaneo-Titiunik-VazquezBare_2020_SAGE}). Part of the popularity of the RD design stems from the idea that the RD treatment assignment resembles the assignment in RCEs, and thus its credibility is similar to the credibility of an actual experiment. The notion of ``as if random'' or ``akin to random'' appears frequently in discussions of RD designs, which suggests that any general discussion surrounding natural experiments should apply to RD designs in particular.

A RD design is a study in which all units receive a score (also known as running variable), and a treatment is allocated according to a specific rule that depends on the unit's score and a known cutoff. In the simplest, binary treatment case, the rule assigns the treatment condition to units whose score is above the cutoff and assigns the control condition to units whose score is below it. Letting $R_i$ be the score for units $i=1,2,\ldots,n$, and $r_0$ be the cutoff, each unit's treatment assignment is $T_i = \I(R_i \geq r_0)$. This rule implies that, conditional on $R$, the treatment assignment is deterministic, since $\P(T_i=1| R_i \geq r_0)=1$ and $\P(T_i=1| R_i < r_0)=0$. All RD designs rely on this discontinuous change in the probability of treatment assignment to study the effect of the treatment at the cutoff, under the assumption that this probability is the only relevant feature of the data generating process that changes discontinuously at the cutoff---or, more precisely, under the assumption that the distribution (or expectation) of the units' potential outcomes is continuous at the cutoff. 

A canonical RD example, first introduced by \cite{Lee_2008_JoE}, is one in which the treatment of interest is winning an election, and the score is the vote share obtained by a political party. Under plurality rules with only two candidates, the party wins the election if it obtains 50\% of the vote or more, and it loses otherwise. Although districts where the party wins will not in general be comparable to districts where the party loses, one interpretation of the RD design poses that in districts where the election is very close, chance plays a role in deciding the ultimate winner.

Some scholars have claimed that the RD treatment assignment rule induces variation in the treatment assignment that is as good as the variation induced by a randomized controlled experiment, elevating RD designs above most other observational studies. The analogy between RD designs and randomized experiments has been invoked frequently to justify the classification of the RD design as an almost-experiment and its treatment assignment as ``as if random''. \citet[p.~7]{Dinardo2016-Palgrave} observes that ``if we focus our attention on the difference in outcomes between `near winners'  and `near losers' such a contrast is formally equivalent to a randomized controlled trial if there is at least some ‘random’ component to the vote share.'' \citet[p.~676]{Lee_2008_JoE} argues that ``causal inferences from RD designs
can sometimes be as credible as those drawn from a randomized experiment'', while \cite{Lee-Lemieux_2010_JEL} call RD designs the ``close cousins'' of randomized experiments. 

These analogies between RD designs and randomized experiments are based on the role of unpredictability in the final treatment assignment. \cite{Dunning2012-Book} sees unpredictability as the source of comparability, asserting that  ``given the role of unpredictability and luck in exam performance, students just above and below the key threshold should be very similar, on average.'' \cite{Lee_2008_JoE} also views uncertainty as the source of comparability, asserting that ``Even on the day of an election, there is inherent uncertainty about the precise and final vote count. In light of this uncertainty, the local independence result predicts that the districts where a party’s candidate just barely won an election (...) are likely to be comparable in all other ways to districts where the
party's candidate just barely lost the election'' \cite[][p.676-77]{Lee_2008_JoE}.

The RD design fits the definition of a natural experiment that I introduced above. Its assignment mechanism is typically neither designed nor controlled by the researcher. Moreover, although it seems that the RD treatment rule $T = \I(R \geq r_0)$ makes the assignment mechanism fully known, it is only known conditional on $R$. Given a unit's score value, the researcher knows whether the probability of being assigned to treatment was zero or one. However, the researcher fundamentally ignores the probability distribution of the score $R$, which implies that, in any window around the cutoff, certain types of individuals could have been more likely than others to receive a score above the cutoff. If types correlate with potential outcomes, then units barely above and barely below the cutoff will not be comparable unless we condition on type. \cite{Sekhon-Titiunik_2017_AIE} discuss this point at length, and show that random assignment of the RD score in a neighborhood of the cutoff does not imply that the potential outcomes and the treatment are statistically independent, nor that the potential outcomes are unrelated to the score in this neighborhood.

This distinction is analogous to the distinction between probabilistic and unconfounded assignment. The element of chance contained in the ultimate value of the score that a unit receives implies that the assignment mechanism is probabilistic. Consider a RD design where a scholarship is given to students whose grade in an exam is above a known threshold. Even good students can see their exam performance adversely affected by ambient noise, unexpected illnesses, or unreasonably hard questions. This means the there is an element of chance in the ultimate grade that any student receives. This element of chance, in combination with the RD rule, implies that a student's placement above or below the cutoff is a random variable. Its probability distribution, however, is fundamentally unknown to the researcher.

Observing the scores assigned to the units in a RD design is analogous to observing the treatment status of each unit in an experiment where the probability of treatment assignment of each unit is hidden from or unknown to the researcher. This means that if we adopt a local randomization approach to RD designs \citep{Cattaneo-Frandsen-Titiunik_2015_JCI,Cattaneo-Titiunik-VazquezBare_2017_JPAM,Cattaneo-Idrobo-Titiunik-Book-2020}, where we focus on a window or neighborhood around the cutoff and use units whose scores are below the cutoff as a comparison group for treated units whose scores are above it, it is natural to imagine that treated units with $R_i = r_0 +\epsilon$ could have instead received a score $R_i = r_0 -\epsilon$ and thus could have been assigned to the control group. It therefore seems plausible to assume that the treatment assignment in a small window around the cutoff is probabilistic, and it is probabilistic by virtue of the unpredictable components of $R$, in combination with the external RD rule $T = \I(R \geq r_0)$. This implies that the RD design satisfies the definition of natural experiment that I have proposed.

My conclusion concurs with \citeauthor{Dinardo2016-Palgrave}'s (\citeyear{Dinardo2016-Palgrave}) and \citeauthor{Dunning2012-Book}'s (\citeyear{Dunning2012-Book}) characterization of the RD design as a natural experiment, but for different reasons. While these authors see the RD design as akin to an experiment, my understanding of the RD as a natural experiment stems from its status as an observational study where an external rule justifies the assumption that the treatment assignment is probabilistic. Understanding RD designs as natural experiments in the sense of definition NE separates the notion of chance from the notion of comparability: the probabilistic nature of the RD assignment implies neither that the RD assignment mechanism is known, nor that it is equiprobable.

 \section{Advantages of Natural Experiments over Traditional Observational Studies}

A natural experiment is fundamentally different from a randomized controlled experiment because its treatment assignment mechanism is unknown to the researcher. For this reason, in the hierarchy of credibility of research designs for program evaluation, natural experiments rank below randomized controlled experiments.\footnote{\cite{DeatonCartwright2018-SSM} \citep[and see also][]{Deaton2010-JEL,Deaton2020-Chapter} reject the idea that research designs can be ranked in terms of credibility. In response, \cite{Imbens2010-JEL} argues that such a ranking is possible in a ceteris paribus sense \citep[see also][]{Imbens2018-SSM}.} At the same time, the most convincing natural experiments rank above other observational studies where the assignment mechanism is not known to depend on a verifiable external factor. The reason is that natural experiments, by virtue of the assignment's dependence on this external factor, offer clear guidelines to distinguish a pre-treatment from a post-treatment period. Moreover, in some cases, the external factor in natural experiments offers a plausible claim of unconfoundedness. 

I shall refer to an observational study where no external factor is known to affect treatment assignment as a \textit{traditional} observational study. As an example of such a study, I consider the influential analysis of the determinants of political participation by \cite{BradyVerbaSchlozman1995-APSR}. These authors propose a resource theory of political participation that expands the traditional socio-economic status (SES) model which focused on income and education as determinants of political participation. Their expanded model is centered in three types of resources: time, money, and civic skills. Their hypothesis is that the amount of each of these three resources available to an individual has a positive effect on that individual's political participation.

 The data come from a representative telephone survey of the United States' adult population that collected self-reported data on respondents' political and civic participation, and also demographic and economic characteristics. Both the outcome (political participation) and the treatment of interest (time, money, and civic skills) are measured with data from this survey. In particular, money resources are measured as self-reported family income; civic skills are measured with educational attainment questions, a vocabulary test, and self-reported participation in nonpolitical organizations such as churches and schools; and time is measured as the hours left in an average day after subtracting time spent sleeping, working, studying, and doing household work.
 
A comparison of this traditional observational study with the natural experiment by \cite{Lassen2005-AJPS} offers important lessons. The assignment mechanism is unknown in both cases. Similarly to the \cite{Lassen2005-AJPS} study, where the probability of each possible allocation of districts to the decentralization condition is unknown, in the  \cite{BradyVerbaSchlozman1995-APSR} study we ignore the probability that each individual will receive a given endowment of money, education, language ability, and free time. There is, however, a fundamental difference. In \cite{Lassen2005-AJPS}, the allocation of districts to the decentralization intervention was the result of a governmental policy. This policy was decided by a third-party, not by  the districts themselves (though we cannot rule out that districts had some influence in determining their own assignment). Moreover, the external mechanism that decided the allocation of districts has a time stamp and is verifiable.

These two features apply to natural experiments generally, and translate into two concrete advantages over traditional observational studies. The time stamp allows the researcher to identify a pre-treatment period, and distinguish it from the post-treatment period. And the verifiability of the external mechanism can, in some cases, justify an uncounfoundedness assumption. I discuss both issues below.

\subsection{Pre-treatment period and falsification}

  In a natural experiment, the assignment mechanism depends on an external factor. As argued at length above, knowledge of this external factor is not sufficient to fully know the probability distribution of the assignment. However, because the occurrence of the external factor is a necessary condition for the treatment to be assigned, the time period when the external event occurs serves as a natural delimiter. Unlike traditional observational studies, natural experiments allow the researcher to establish objectively the time period when treatment assignment occurs, because she can record when the external intervention was initiated.
  
This treatment assignment time stamp is crucial for falsification purposes. Once the researcher collects information about the moment when the treatment was given to the units, the periods before and after the treatment assignment are easily established---the period before the treatment is commonly referred to as the \textit{pre-treatment period}. An important falsification analysis is available if researchers can collect information on a set of covariates $\mathbf{X}$ measured during the pre-treatment period. By virtue of having been measured in this period, these variables will be determined before the treatment is assigned and thus the effect of treatment on them is zero by construction. Thus, the variables $\mathbf{X}$ can be used to implement a falsification analysis that is common in the analysis of randomized experiments: by analyzing whether the treatment has in fact no effect on the covariates, researchers can offer empirical evidence regarding the comparability of treated and control groups. 

As in randomized experiments, the usefulness of this so-called ``covariate balance'' analysis depends on the type of variables that are included in  $\mathbf{X}$. The most convincing falsification analysis will be one where these variables that are strongly correlated with both the outcome and the factors that affect the propensity to receive the the externally-assigned treatment.\footnote{For example, in the \cite{Lassen2005-AJPS} study, one could analyze the share of the population that is college-educated, which is known to correlate with voter turnout (the outcome of interest), and is also correlated with socio-economic indicators such as income and poverty that might make decentralization (the treatment) more or less desirable.} On this aspect, natural experiments do not differ much from RCEs and RTPEs.

However, there is one crucial difference. The correct implementation of a covariate falsification analysis depends on the assignment mechanism, which in a natural experiment is unknown. When the assignment mechanism is equiprobable, the distribution of $\mathbf{X}$ is expected to be the same when the entire control group is compared to the entire treatment group, and thus the falsification test can be implemented with unadjusted covariate balance tests that compare all treated units versus all control units.\footnote{An equiprobable assignment is one in which every unit has the same probability of receiving treatment, but not necessarily one in which this probability is equal to 50\%. As long as this probability is constant for all units, the distribution of covariates in the treatment and control group will be the same.} However, if the treatment assignment probabilities are different for different subgroup of units, the proper implementation of a covariate balance test requires to weight or stratify the analysis based on these probabilities. In natural experiments, however, these probabilities are unknown, so such adjustment is unavailable.

This suggests that an unadjusted covariate balance test is a useful tool to establish the plausibility of the equiprobable assignment assumption. For implementation, researchers should assume that the assignment mechanism is equiprobable and test the hypothesis that the unadjusted distribution of $\mathbf{X}$ is  equal in the treatment and control groups. If the hypothesis that the treated and control covariate distributions are equal is rejected, then the assumption of equiprobable assignment is unsupported by the data. This is an important first step to gain a deeper understanding of the assignment mechanism.

The implementation of this falsification analysis is straightforward in the \cite{Lassen2005-AJPS} study. The decentralization intervention occurred in 1995 when the Copenhagen Municipality Structural Commission selected the districts that would be decentralized. Thus, all district-level variables collected before 1995 are pre-treatment and can be used in a falsification analysis. This could include census counts, economic indicators, etc. In contrast, in the \cite{BradyVerbaSchlozman1995-APSR}, the pre-treatment period is impossible to identify with certainty because it is unclear when the treatments of time, money and civic skills are in fact assigned. For example, if an individual reports high levels of civic skills as measured by a vocabulary test, what exactly is the period before this skill was developed? We know that language skills are susceptible to stimulation since an early age, and toddlers and even infants who are exposed to rich language environments have stronger language skills. We cannot rule out that people with high vocabulary skills have been exposed to this treatment since early childhood. A similar argument can be applied to the money and time treatments. This implies that a pre-treatment period is unavailable and all covariates are in fact post-treatment covariates. Therefore, there are no covariates available with which to implement a falsification analysis.

\subsection{Verifiability of externality and unconfoundedness}
When the empirical evidence shows that the distribution of relevant predetermined covariates differs between the treatment and the control group, the assumption of equiprobable assignment is implausible. This means that the data does not support the assumption that all units were assigned to the treatment condition with the same probability. Without additional knowledge, it is not possible to identify causal treatment effects in a design-based fashion. 

However, the most convincing natural experiments might offer a reasonable justification for the assumption that the assignment is unconfounded given some observable predetermined covariates. The credibility of this justification is based directly on the externality of the treatment assignment that characterizes natural experiments. As I have defined it, a natural experiment is a setting in which the treatment assignment mechanism is known to be probabilistic by virtue of depending on an external factor. In some natural experiments, the researcher has enough information about the variables on which the external intervention depended. In these cases, the researcher might credibly assume that, after these variables are conditioned on, the probability of treatment assignment is not a function of the units' potential outcomes. The credibility of such an assumption should be judged on a case-by-case basis.

For example, in the \cite{Lassen2005-AJPS} study, the exact functional form of the assignment mechanism is unknown, but a qualitative investigation of the decision-making process revealed that the decision of which districts to decentralize was based on the districts' population and socioeconomic development, with the explicit goal of ensuring that the decentralized districts were as diverse as the total population of districts in terms of these covariates. This feature of the assignment, which is directly verifiable with qualitative information issued by the Copenhagen Municipal Commission, can be used as the basis for the unconfoundedness assumption that the probability of decentralization is unrelated to the districts' potential outcomes after conditioning on population and socioeconomic development.

Note an important difference between the unconfoundedness assumption and the equiprobable assignment assumption: the latter is empirically testable, but the former is not. Because covariate balance is an implication of an equiprobable assignment mechanism, we can use covariate balance tests to falsify the assumption that the assignment mechanism is equiprobable. However, in the absence of additional assumptions, the unconfoundedness assumption is fundamentally untestable. This means that a justification for it has to rely more heavily on the qualitative information about the assignment mechanism, and stands on weaker evidentiary ground.

The assumption of unconfounded assignment is always strong, but on this respect natural experiments have an advantage over traditional observational studies: the dependence of the assignment mechanism on external factors is verifiable. In the most convincing natural experiments, researchers are able to verify that the process that governed the treatment assignment depended on external factors, and prior scientific knowledge coupled with qualitative and/or qualitative data suggests that treatment assignment should be unrelated to potential outcomes conditional on those factors. If the researcher is able to collect information on those same factors and condition on them in the analysis, then the usual tools of program evaluation based on unconfoundedness---parametric adjustment models, propensity score analyses, matching estimators, etc.---are available for analysis.

Thus, in a convincing natural experiment, the researcher uses the available information on the  external assignment mechanism as a plausible basis to invoke an unconfoundedness assumption. This is unavailable in a traditional observational study, where there is usually no objective basis to claim that unconfoundedness holds for any set of covariates, given that we fundamentally ignore how (and when) the treatment was assigned. For example, in the \cite{BradyVerbaSchlozman1995-APSR} study, what covariates should we condition on before we can assume that people with high levels of money, time and civic resources are comparable to people who have low levels of those resources? \citeauthor{BradyVerbaSchlozman1995-APSR} condition on citizenship status because they reasonably assume that it is a  ``prerequisite for voting and might affect other kinds of participation as well''. But, even putting aside the concerns about establishing the pre-treatment period, we can imagine many other factors such as geographic location, number of children, parents' education, etc., that may affect both the propensity to participate in politics and the amount of time, money, and resources available to an individual. There is no objective information to guide the choice of the conditioning set.

The decision to participate in politics, since it is made privately and is entirely under the control of each individual, is less transparent to the researcher than the decision to decentralize districts in Copenhagen. Unlike the Copenhagen Municipal Commissions, which published a report on the decentralization process, individual citizens do not write reports detailing the process by which they arrived at the decision to participate in politics. This greater transparency about the assignment mechanism, and the separation between the units receiving the treatment and those assigning it, can imbue some natural experiments with a stronger research design and more objective basis to invoke the necessary identification assumptions. 

My choice of \textit{can} in the prior sentence is deliberate, and should not be replaced by \textit{do}. I do not mean to claim that the ``worst'' natural experiment is always preferable to the ``best'' traditional observational study. Some natural experiments blatantly violate the equiprobable assignment assumption and provide a very weak basis for assuming unconfoundedness. Some traditional observational studies are carefully conducted and contribute to our scientific knowledge. My claims about a credibility hierarchy are made in the ceteris paribus spirit articulated by \cite{Imbens2010-JEL}---in a given study, it is preferable to have a verifiable conditioning set and a clear time stamp attached to the treatment assignment, and no researcher would willingly give up such information.

\section{Recommendations for Practice}

The preceding discussion suggests some general recommendations for empirical researchers who wish to estimate and interpret causal effects based on natural experiments.

\noindent\textit{ (1) Is the assignment probabilistic?} The first step is to establish whether the assumption of a probabilistic assignment is met for the universe of units that the researcher wishes to analyze. As I have defined it, a crucial feature of a natural experiment is that its assignment mechanism is probabilistic by virtue of an external event that is outside of the units' direct control. The researcher should establish whether it is in fact the case that all units to be included in the study had a probability of receiving treatment strictly between zero and one. If some units were certain to either be affected or not affected by the intervention, they should be excluded from the study, as the usual causal parameters will not be identifiable. If some units are excluded, the researcher should redefine the parameter of interest and clarify in the analysis that the reported effects are estimating the effect of the intervention only for units whose probability of being treated was neither zero nor one. The researcher should carefully characterize this new parameter.

The caveat is that the assumption of probabilistic treatment assignment is not directly verifiable or testable, because untreated units could be untreated either because their ex-ante treatment assignment probability is zero or because it is positive but the realization of the assignment is the control condition. For this reason, researchers should use prior scientific knowledge and/or qualitative and quantitative information regarding the external process that assigned the treatment to justify the probabilistic assignment assumption.

\noindent\textit{(2) Is the assignment equiprobable?} The second step is to assume that the assignment mechanism is equiprobable and test the hypothesis that the distribution of relevant pre-treatment covariates is equal in the treatment and control groups. This falsification analysis starts by selecting a group of relevant pre-treatment covariates $\mathbf{X}$ and testing the null hypothesis that the means and other features of the distribution of these covariates are the same in treated and control groups. If the hypothesis of covariate balance is not rejected, the analysis can proceed under the equiprobable assignment assumption, using standard tools from the analysis of randomized experiments \citep[e.g.,][]{AtheyImbens2017-HandbookFE,GerberGreen2012-Book,ImbensRubin2015-Book}---with the caveat that in natural experiments, unlike in RCEs or RTPEs, this assumption is not known to be true and its credibility might be disputed by other analyses. If the hypothesis of covariate balance is rejected, then the assumption of equiprobable assignment is unsupported by the data. Of course, researchers should ensure that their tests have enough statistical power to avoid mistakenly interpreting the failure to reject a false null hypothesis of covariate balance as supportive of the equiprobable assignment assumption.

\noindent\textit{(3) Is the assignment unconfounded?} An assignment mechanism that is not equiprobable could still be unconfounded. When the data do not support the assumption of equiprobable assignment, researchers should explore whether it is plausible to assume that there exists a covariate-based adjustment that renders treated and control groups comparable.  In this second stage of falsification, researchers can use the external assignment mechanism of the natural experiment to offer a plausible basis to adopt an unconfoundedness assumption. This justification should be based on objective and verifiable information about the treatment assignment mechanism that identifies a set of covariates that were explicitly used in the assignment, as in the \citeauthor{Lassen2005-AJPS} study. Assuming that the researcher has access to these covariates, the analysis can proceed under the assumption of unconfounded given these covariates using standard estimation and inference methods from the unconfoundedness toolkit \cite[e.g.,][]{Abadie-Cattaneo_2018_ARE, ImbensRubin2015-Book}---again, with the caveat that this assumption is not known to be true and might be disputed by later analyses.

\noindent\textit{(4) Is the natural experiment of substantive interest?}  In most natural experiments, the treatment that is assigned is not exactly the treatment that a researcher would have assigned if she had been in charge of the execution of the study. This leads to very important and often difficult issues of interpretation. Even if all the required identification assumptions are satisfied, the treatment effect that is identifiable by the design may not be the effect of scientific interest.

\cite{SekhonTitiunik2012-APSR} illustrate this point with a redistricting natural experiment. Several researchers have used the periodic re-drawing of district boundaries to study the incumbency advantage, comparing the vote share received by the same incumbent legislator in areas that are new to her district versus areas that have been part of the district for a long time. Even if precincts are randomly moved to new districts  according to a known probability distribution, this assignment would never achieve comparability between new and old voter areas in terms of their prior history (e.g. party or race of prior incumbent), because new voters are coming from a different incumbent by construction. In terms of the prior discussion, this occurs because the probability of old voters of being selected as new voters is zero, and thus the overall assignment is not probabilistic. The natural experiment externally introduces variation in the voters that an incumbent receives in her/his district. Whether this variation is useful to study the incumbency advantage of interest to scholars of American politics is a separate matter. Such issues of interpretation should be at the forefront of any analysis based on natural experiments. 

\section{Conclusion}

The literature has offered several definitions of a natural experiment, not necessarily consistent with one another. I sought to partly resolve the ambiguity by going back to the definition of a randomized controlled experiment, and contrasting the canonical natural experiment to it. As I have defined it, a natural experiment is a study in which the treatment assignment mechanism is neither designed nor implemented by the researcher, is unknown to the researcher, and is probabilistic by means of an external event or intervention that is outside of the control of the units who are the subject of the intervention. 

In order to arrive at this definition, I have emphasized several conceptual distinctions. A central conclusion is that a randomized controlled  experiment's defining feature is not that the treatment assignment is random, in the sense of being a random variable with some distribution, because this would imply that all interventions, programs, and individual decisions ever taken are randomized experiments. That a citizen's decision to vote is a random variable does not imply that a comparison of voters and nonvoters is a randomized experiment. The key is not that the treatment must have a distribution (all random variables do) but rather that the experimenter must know what this distribution is. The power of a randomized controlled experiment is therefore not only in the randomization itself, but also in the knowledge and properties of the assignment distribution that the randomization implies. In a randomized controlled experiment, the unconfoundedness assumption guaranteed by the physical randomization device is as crucial as the ex-ante unpredictability of each individual's treatment assignment. In contrast, natural experiments retain the unpredictability but discard knowledge of the assignment mechanism and the unconfoundedness guarantees. 

Because natural experiments have, by definition, a treatment assignment mechanism that is unknown to the researcher, they rank---everything else equal---unambiguously below randomized controlled experiments in terms of credibility and reproducibility. Nonetheless, natural experiments offer two important advantages over traditional observational studies. First, by defining the moment when the intervention of interest occurs, they clearly demark a pre-treatment period, which is essential to falsify the assumption of equiprobable assignment and also to condition on covariates in a valid way. Second, in cases where the equiprobable assignment assumption does not hold, the best natural experiments offer a plausible and verifiable justification for an unconfoundedness assumption. Both the time stamp that delimits pre and post treatment periods and the objective justification for the unconfoundedness assumption are often lacking in traditional observational studies.

\newpage 
\bibliographystyle{apsr_fs.bst}
\bibliography{Titiunik-NaturalExperiments.bib}

\end{document}